\documentstyle[psfig,conf_iap,11pt]{article}
\begin{document}
{\small 
\heading{What causes the Ly$\alpha$ forest, clouds or large-scale velocity fields ?} 
\par\medskip\noindent
\author{Wilhelm H. Kegel and Sergei A. Levshakov}
\address{Institut f\"ur Theoretische Physik der Universit\"at Frankfurt am Main, 
Postfach 11 19 32, 60054 Frankfurt/Main 11, Germany }
\address{Department of Theoretical Astrophysics, A. F. Ioffe Physico-Technical 
Institute, 194021 St.Petersburg, Russia}

If an additional
{\it large scale} hydrodynamic velocity field is superposed on the general Hubble flow, 
the peculiar velocities have the effect that spatially separated volume elements along a given line of
sight may contribute to the absorption at the same value of $\lambda$. This effect leads to a 
{\it `line-like'} absorption structure rather than to a smooth GP-depression. 
{\bf From this it follows 
that at least part of the Ly$\alpha$ forest may be explained without invoking density fluctuations
(clouds). Thus, there is no clear observational
distinction between intergalactic clouds and the diffuse medium inbetween}.

We have investigated this effect by means of Monte Carlo simulations \cite{LK}.
In Fig.~1 we present results
based on a model in which we assumed the diffuse intergalactic medium to be (locally) homogeneous
having at z = 3 the parameters $n({\rm HI}) = 4\times10^{-11} \  {\rm cm}^{-3}$ and $T_{\rm kin} = 10^4 $K.
For the large scale stochastic velocity field we assumed a rms velocity 
of $\sigma_{t} = 300\ {\rm km\ s}^{-1}$,
and a correlation length $l = 6$ Mpc. In calculating the expansion rate we assumed $q_0 = 0.5$ and
${\rm H}_0 = 70$ km s$^{-1}$ Mpc$^{-1}$. The value of $n({\rm HI})$ was estimated from the
continuum depression D$_A$ at $z = 3$ observed in low resolution spectra, $T_{\rm kin}$ 
corresponds to the width of the narrowest lines in the  Ly$\alpha$ forest, $\sigma_{t}$ is half
the value found for the peculiar velocities in the local universe \cite{RS}, 
and the value of  $l$ corresponds to the expected size of voids at $z = 3$.  
We performed similar calculations for He~II Ly$\alpha$ (Fig.~2).

Assuming a photoionizing rate of $\Gamma_{\rm HI} = 10^{-12}$ s$^{-1}$ \cite{HM}
one derives from $n({\rm HI})$ a total baryon density of the order of
$10^{-5}$ cm$^{-3}$ at z = 3 corresponding to $\Omega_{0, {\rm IGM}} \simeq 0.03$, i.e.
it is about 3 per cent of the mass needed to close the Universe. This result indicates
that most of the baryonic matter is found in the IGM inbetween clouds.

\acknowledgements{This work was supported in part by the RFBR grant
No. 96-02-16905-a and by the Deutsche Forschungsgemeinschaft. }

\begin{figure}
\vspace{2.0cm}
\centerline{\vbox{
\psfig{figure=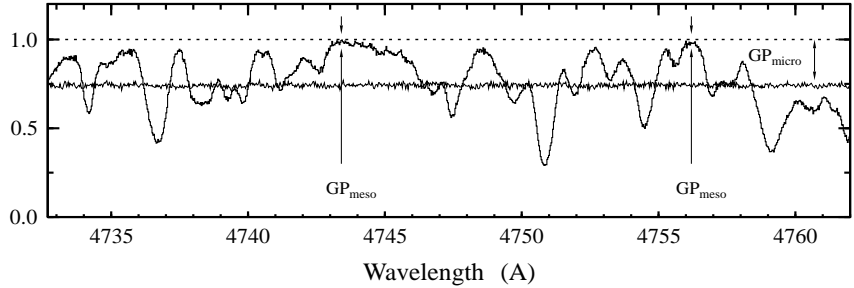,height=2.5cm,width=13.0cm}
}}
\vspace{-1cm}
\caption[]{Monte Carlo simulation of the HI Ly$\alpha$ forest for a given
velocity field realization at $z = 3$ (S/N = 100, FWHM = 8 km s$^{-1}$).
The horizontal `noisy' line shows a result for zero correlation coefficient,
representing a classical GP-trough (marked by GP$_{micro}$). Commonly 
the diffuse intergalactic
$n({\rm HI})$ is estimated by measuring the
intensity distribution near the apparent continuum, these regions are
labeled by GP$_{meso}$.
}
\end{figure}
%
\begin{figure}
\vspace{2.0cm}
\centerline{\vbox{
\psfig{figure=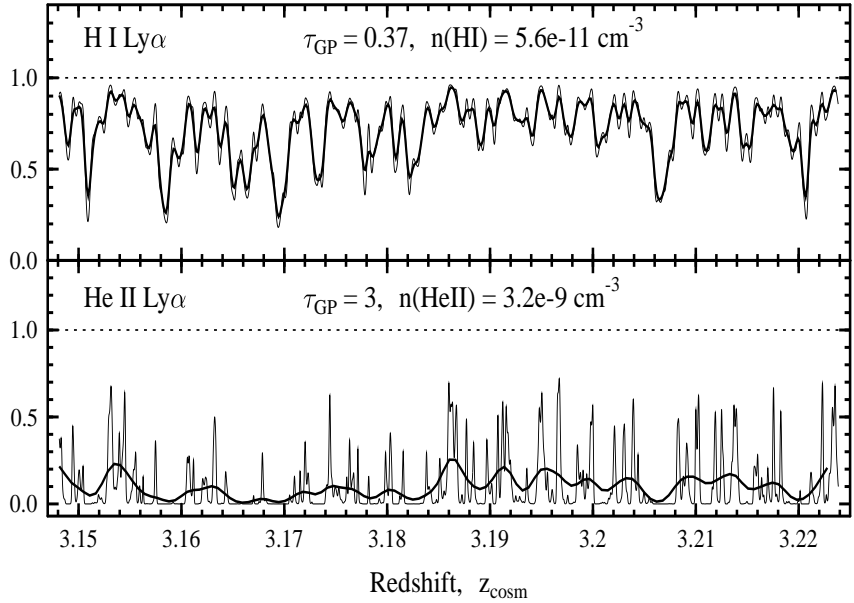,height=3.0cm,width=13.0cm}
}}
\vspace{2.7cm}
\caption[]{Monte Carlo simulation of the HI (upper panel) and the corresponding
HeII (lower panel) Ly$\alpha$ forest at $z = 3.2$ (thin lines) calculated with
S/N = 50. Also shown (thick lines) are the same spectra convolved with a spectrograph
function of a 0.6 \AA\ width.
}
\end{figure}

\vfill
}
\end{document}